\documentclass{aa}
\usepackage{epsfig,times}
\def\mincir{\raise -2.truept\hbox{\rlap{\hbox{$\sim$}}\raise5.truept \hbox{$<$}\ }}
\def\mincireq{\hbox{\raise0.5ex\hbox{$<\lower1.06ex\hbox{$\kern-1.07em{\sim}$}$}}}
\def\magcir{\raise-2.truept\hbox{\rlap{\hbox{$\sim$}}\raise5.truept \hbox{$>$}\ }}
\def\gr{\kern 2pt\hbox{}^\circ{\kern -2pt K}} 

\def\_{\thinspace}

\def\ea{\ et al. \,}

\def\be{\begin{equation}}
\def\ee{\end{equation}}

\begin{document}

\title{VHE emission from M\,82}

\author{Massimo Persic\inst{1}, 
        Yoel Rephaeli\inst{2,} \inst{3},
	and Yinon Arieli \inst{2,} \inst{3} }
            
\offprints{M.P.; e-mail: {\tt persic@oats.inaf.it}}

\institute{
INAF/Osservatorio Astronomico di Trieste and INFN-Trieste, via 
G.B.Tiepolo 11, 34143 Trieste, Italy
	\and
School of Physics \& Astronomy, Tel Aviv University, Tel Aviv 69978, Israel
        \and
Center for Astrophysics and Space Sciences, University of California at 
San Diego, La Jolla, CA 92093, USA
           }
\date{Received ..................; accepted ...................}

\abstract{

Spurred by the improved measurement sensitivity in the very-high-energy 
(VHE: $\geq$100 GeV) $\gamma$-ray band, we assess the feasibility of 
detection of the nearby starburst galaxy M\,82. VHE emission is expected 
to be predominantly from the decay of neutral pions which are produced in 
energetic proton interactions with ambient protons. An estimate of VHE 
emission from this process is obtained by an approximate, semi-quantitative 
calculation, and also by a detailed numerical treatment based on a 
convection-diffusion model for energetic electron and proton propagation 
and energy losses. All relevant hadronic and leptonic processes are 
considered, gauged by the measured synchrotron radio emission from the 
inner disk region. We estimate an integrated flux $f($$\geq$100\,GeV)$
\sim$2$\times$10$^{-12}$ cm$^{-2}$s$^{-1}$, possibly detectable 
by the current northern-hemisphere imaging air Cherenkov telescopes, MAGIC 
and VERITAS, and a good candidate for detection with the upcoming MAGIC~II 
telescope. We also estimate $f($$\geq$100\,MeV)$\sim$10$^{-8}$ cm$^{-2
}$s$^{-1}$, a level of emission that can be detected by {\it GLAST/LAT} 
based on the projected sensitivity for a one year observation.

}

\maketitle
\markboth{Persic et al.: VHE emission from M\,82}{}

\section{Introduction}

In the central regions of starburst galaxies (SBGs) high level of star 
formation (SF) activity powers the emission of radiation directly by 
supernovae (SN) and massive stars, and indirectly by SN shock heating 
of interstellar gas and dust, as well as from radiative processes 
involving electrons and protons that are accelerated by SN shocks. 
Kinetic energy of the non-thermally distributed electrons and protons 
is partly channeled into radio to the high-energy $\gamma$-ray emission. 
In particular, radiation in the very-high-energy (VHE: $\geq$100 GeV) 
$\gamma$-ray region is emitted predominantly in $\pi^0$ decay, following 
the production of the pions in energetic proton interactions with protons 
in the gas, and by relativistic electrons traversing (relatively) strong 
magnetic field and scattering off the intense radiation field in the 
galactic disk.

The level of high-energy phenomena in intensely active galaxies makes 
AGN obvious targets of searches for VHE emission, and indeed some 
15 blazars have already been detected by H.E.S.S. and MAGIC (e.g., 
Persic \& De Angelis 2008). Various issues in the study of non-thermal  
phenomena in `normal' and starburst galaxies are no less interesting and 
important. Valuable insight can be gained on the origin and propagation 
modes of relativistic electrons and protons by probing their radiation 
yields across the electromagnetic spectrum. Knowledge of the low energy 
densities of these particles is important also for quantifying their 
impact on interstellar gas.

At a distance $d$$=$3.6 Mpc (Freedman et al. 1994), the SB galaxy M\,82 is 
a good candidate for an initial search for TeV emission from a 
galaxy whose emission is not AGN-dominated. 
Its far-infrared (FIR) and X-ray luminosities imply high SF rates (e.g., 
Persic \& Rephaeli 2007), and a correspondingly high SN rate of 0.3 yr$^{-1}$ 
(Rieke et al. 1980). In most respects, levels of activity in M\,82 surpass 
those in the other nearby starburst galaxy, NGC\,253, making it a very good 
target for TeV observations, also in view of its large reservoir of diffuse 
gas ($\sim$10$^9$$M_\odot$; Casasola et al. 2004). 

In order to realistically estimate non-thermal emission from cosmic-ray 
electrons and protons, we start with the initial spectrum that characterizes 
their momentum distribution at the acceleration sites, and follow the 
evolution of the spectrum as the particles diffuse and are convected from 
the inner SB region to the outer disk (and halo). All the relevant leptonic 
and hadronic interactions are included in a numerical program we have 
developed for this purpose. Using this program we can determine the particle 
spatial and energy distributions, and their respective radiative yield across 
the galaxy. For a more intuitive approach, and in order to compare with 
previous semi-quantitative estimates, we also use a simple analytical model 
of the diffuse hadronic emission of M\,82, that ties the predicted 
$\gamma$-ray flux to the measured on-site synchrotron radio flux by assuming 
energetic particles and magnetic field to be in energy equipartition. 

In Section 2 we briefly review the basic features of predicted particle 
source spectra, the relation between the cosmic-ray electron and proton 
components, and key aspects of their energy-loss processes and propagation 
in galaxies. We then summarize the main features of our numerical program 
which is used to determine the particle spectro-spatial distribution. 
In Section 3 we present a specific estimate of the hadronic emission from 
M\,82 based on our approximate analytical treatment. Results of our detailed 
numerical analysis are presented in Section 4, and discussed in Section 5.

\section{Particle spectra and energy-loss processes}

\subsection{Particle source spectra}

SN shock wave sweeps and compresses circumstellar material with a 
compression ratio $s$$\equiv$$(\gamma_c$$+$1)/[$\gamma_c$$-$1$+$2$\, 
(\beta/M)^2]$, where $\gamma_c$$\equiv$$C_P/C_V$$=$5/3 is the 
specific heat ratio, $\beta$$\equiv$$(c_s/V_A)$ is the speed of sound 
($c_s$) in units of the Alfv\'en velocity ($V_A$), and $M \equiv 
u_1/V_A$ is the Alfv\'enic Mach number of the shock, with $u_1$ 
denoting the upstream velocity of the gas. Particle acceleration by 
the shock modifies the momentum distribution to a power-law whose index 
depends on $M$. When particles with initial momentum $p_0$ are overtaken 
by the shock, the resulting downstream phase space distribution function 
is $f(p) \propto\,{\bf p}^{-\eta}$ (for $p$$>$$p_0$) where 
$\eta$$=$$3s/$$(s$$-$1)$=$$3M^{2}$$(\gamma$$+$1)/2$(M^2$$-$$1)$ 
(Krymsky 1977, Axford et al. 1977, Bell 1978, and Blandford \& Ostriker 1978). 
The ensuing number density of accelerated particles per momentum interval 
$dp$ is $N(p) \propto f(p)p^{2} \propto p^{-q}$ with 
$q$$=$$\eta$$-$2. It can be readily 
seen that in the relativistic regime the particle density per unit energy 
interval has the same index, $q$. The strong shock limit has $\eta$$=$4, 
i.e. $q$$=$2, a well-known result for the first-order Fermi acceleration 
mechanism (e.g., Protheroe \& Clay 2004). 

Particles can be accelerated to very high energies ($\geq$10$^{14}$ 
eV) even by the non-relativistic shocks in SN remnants 
(Aharonian et al. 2007), which are the accelerators of relevance to 
us here. The steepening of the particle power-law distribution to 
$q>2$ (due to energy losses outside the source region) implies that 
-- for our purposes here -- the specific value of the high-energy cutoff 
is irrelevant. For non-relativistic shock 
acceleration 
the lowest kinetic energy can be assumed to be the thermal energy of the 
gas, which is the source of accelerated particles.


It is perhaps useful to outline the following simple -- if physically 
uncertain -- 
prediction (Schlickeiser 2002; Bell 1978) for the proton to electron (p/e) 
ratio: Assume that electrons and protons are accelerated out of a thermal 
($T_0$$\sim${\it few} keV), electrically neutral plasma reservoir to high 
($T$$\geq$$T_0$) kinetic energies according to the same differential 
power-law momentum spectrum, $N_{\rm i}(p)$$=$$N_{\rm i,0} p^{-q}$ with 
i$=$e,p, maintaining charge neutrality.
\footnote{ 
	It may be argued that charge neutrality does {\it not} 
	necessarily require equal numbers of positive and negative 
	charges to be accelerated. Very minute current flows in the 
	thermal plasma can easily compensate for any charge imbalance 
	in the accelerated particles. Indeed, because electrons have 
	much higher losses and different diffusion properties,  
some degree of charge neutrality could develop during propagation 
away from the acceleration sites.} 

Under these assumptions, the p/e number and energy density ratios in the 
source region can be analytically calculated. As function of kinetic energy 
$T$$=$$(\gamma$$-$$1)$$mc^2$, where $\gamma$ is the particle 
Lorentz factor, the p/e number density ratio is found to be 
\begin{eqnarray}
{N_{\rm p}(T) \over N_{\rm e}(T)} 
~\simeq~ 
\biggl({m_{\rm p} \over m_{\rm e}}\biggr)^{q-1 \over 2} ~ 
\biggl({T+m_{\rm p}c^2 \over T+m_{\rm e}c^2}\biggr) ~
\biggl({T+2m_{\rm p}c^2 \over T+2m_{\rm e}c^2}\biggr)^{-{q+1 \over 2}} \, , & &
\label{eq:numratio1}
\end{eqnarray}
with $m_{\rm e}$ and $m_{\rm p}$ the electron and proton masses. This ratio 
assumes the following limiting forms: 
\begin{eqnarray}
{N_{\rm p}(T) \over N_{\rm e}(T)} \simeq \left\{ \begin{array}{lll}

1  & \mbox{$T << m_{\rm e}c^{2}$}\,;\\

[T/(m_{\rm e}c^2)]^{q-1 \over 2} & \mbox{$m_{\rm e}c^2 << T << m_{\rm p}c^2$}\,;\\
 
({m_{\rm p} \over m_{\rm e}})^{q-1 \over 2} & \mbox{$m_{\rm p}c^2 << T$\,.} 
\end{array} \right.
\label{eq:numratio2}
\end{eqnarray}
In particular, the ratio reaches its maximum value, $(m_{\rm p}/m_{\rm e})^{
(q-1)/2}$ (for $q$$>$1), over most of the range of particle energies. 

Similarly, the p/e energy density ratio, 
\begin{eqnarray}
\kappa~\equiv~ {U_{\rm p} \over U_{\rm e}} ~=~ 
{ \int_{T_0}^\infty N_{\rm p}(T) ~T~{\rm d}T \over 
\int_{T_0}^\infty N_{\rm e}(T) ~T~{\rm d}T }\,,  & &
\label{eq:kappa_def}
\end{eqnarray}
can be expressed in terms of the spectral index $q$, 
\begin{eqnarray}
\lefteqn{
\kappa(q) ~=~ }
		\nonumber \\
& {
\biggl[{T_0^2 \over c^2} + 2T_0m_{\rm p} \biggr]^{q-1 \over 2} 
\int_{T_0}^\infty 
T (T+m_{\rm p}c^2)  \biggl[{T^2 \over c^2} + 2Tm_{\rm p} \biggr]^{-{q+1 \over 2}} {\rm d}T
\over 
\biggl[{T_0^2 \over c^2} + 2T_0m_{\rm e} \biggr]^{q-1 \over 2} 
\int_{T_0}^\infty 
T (T+m_{\rm e}c^2)  \biggl[{T^2 \over c^2} + 2Tm_{\rm e} \biggr]^{-{q+1 \over 2}} {\rm d}T
} 
& \,.
\label{eq:kappa_for}
\end{eqnarray}
For example, if $q$$=$2.3, then $U_{\rm p}$/$U_{\rm e}$$\sim$15, and 
$N_{\rm p}$/$N_{\rm e}$$\sim$1.3$\times$10$^2$ at $T$$>>$1 GeV. These 
are approximately the values measured for Galactic cosmic rays (e.g., 
Schlickeiser 2002). 

The spectrum of relativistic electrons is obviously more easily measured 
due to their much more efficient radiative losses. Most readily observed 
is electron synchrotron radio emission, which we will use to infer 
$N_{\rm e}(\gamma)$ in the region where the radio emission is observed. 
To relate this quantity to the source spectrum we need to solve the 
kinetic equation describing the propagation mode and energy losses 
incurred by electrons as they move out from their SN shock regions. 
Using the above predicted proton to electron ratio at their common region, 
and accounting for proton propagation mode and energy losses, we can then 
deduce the proton spectrum from the electron spectrum. Clearly, even if the 
propagation modes of protons to electrons are similar, their losses are very 
different; thus, $N_{\rm p}$ should be inferred from particle number 
conservation.

\subsection{Steady-state particle spectra}

The enhanced SF activity in a starburst galaxy lasts for $O(10^{8})$ yr, which 
is also the timescale over which global SN shock acceleration is 
prevalent. Estimated times for propagation out of the disk and inner 
galactic halo are about an order of magnitude lower for low energy 
particles. High energy particles lose energy on much shorter timescales. 
Thus, particle spectra evolve considerably as they propagate out of 
their acceleration sites. As long as typical acceleration times are 
comparable to or shorter than loss times for sufficiently high energies, 
a steady state can be attained, with particle spectra which are obviously 
steeper than their respective source spectra. Strictly speaking, we 
assume that steady state {\it is} attained and proceed to solve the 
kinetic equation for $N_{i}(\gamma, R,z)$, where $R$ and $z$ are the 
2D spatial radius and the coordinate perpendicular to the galactic 
plane, respectively. 

A full calculation of the particle steady state spectra is necessarily 
extensive as it must include all the important energy loss mechanisms 
and modes of propagation. We have employed the numerical code of 
Arieli \& Rephaeli (Arieli \& Rephaeli, in preparation), which is based 
on a modified version of the GALPROP code (Moskalenko \& Strong 1998, 
Moskalenko \ea 2003), that has been developed for this purpose. The 
code solves the exact Fokker-Planck transport diffusion-convection 
equation (e.g., Lerche \& Schlickeiser 1982) in 3D with given source 
distributions and boundary conditions for electrons and protons. 
Evolution of the particle energy and spatial distribution function is 
based on diffusion, and convection by galactic wind. (We note that some 
of the additional processes included in the original code, such as 
diffusive re-acceleration, nuclear fragmentation, and radioactive decay, 
are not relevant for our purposes here.) Logarithmic coordinates are 
used for high spatial resolution. 

The main focus of this first phase of our work is the estimation of 
the high energy flux of M\,82. An approximate estimate of the high 
energy $\gamma$-ray emission can be obtained by a simple and intuitive 
calculation that is based on sampling the electron energy spectrum 
from the main observable -- the radio synchrotron flux. However, 
because the electron spectrum needs to be extrapolated to energies 
much higher than the $\sim 1-10$ GeV range directly inferred from 
radio measurements, realistic estimates of the TeV emission necessitates 
a detailed (numerical) treatment. To illustrate this, and to provide 
insight on the results from the more accurate numerical calculation, 
we first outline the intuitive calculation.

\subsection{Energy-loss processes}

At energies below few hundred MeV, electrons lose energy mostly by 
Coulomb interactions with gas particles, leading to ionization of neutral 
and charged ions, and electronic excitations in fully ionized gas. At 
higher energies the dominant energy losses are via synchrotron emission, 
and Compton scattering by the FIR and optical radiation fields. 
These well known processes need no elaboration; the level of the ensuing 
emission depends on the mean strength of the magnetic field, $B$, and the 
energy density of the radiation fields, which are specified below.

\subsubsection{Gauging particle spectra by radio emission}

The electron population consists mostly of directly accelerated (primary) 
electrons, which can be represented by a single power-law (in terms of the 
Lorentz factor) 
\begin{eqnarray}
N_{\rm e}(\gamma) ~=~ N_{e,{\rm 0}} ~ \gamma^{-q} & &
\label{eq:el_spectrum}
\end{eqnarray}
for $\gamma_1$$\leq$$\gamma$$\leq$$\gamma_2$. The electron synchrotron flux 
from a spherical region of radius $r_{\rm s}$ with magnetic field $B$, 
located at a distance $d$, is 
\begin{eqnarray}
\lefteqn{
f_\nu ~=~  ~ 5.7 \times 10^{-22}~ { r_{\rm s}^3 \over d^2}  
N_{\rm e,0} ~a(q) ~B^{q+1 \over 2} ~ \times ~}
                \nonumber\\
& & {} ~~~~~~~~~~~~~ \times ~ \bigg({\nu \over 4 \times 10^6}\biggr)^{-{q-1 \over 2}} 
~~~ {\rm erg/(s~ cm^2 Hz)}
\label{eq:synchr}
\end{eqnarray}
where $a(q)$ is defined and tabulated in, e.g., Tucker (1975). 

Setting $\nu$$=$1$\,$GHz and 
$\psi$$\equiv$
$( {r_{\rm s} \over 0.1 \,{\rm kpc}})^{-3}$ 
$( {d \over {\rm Mpc}})^{2}$ 
$( {f_{1\, {\rm GHz}} \over {\rm Jy}})$, 
from Eq.(\ref{eq:synchr}) the normalization of the electron spectrum is 
\begin{eqnarray}
N_{\rm e,0} ~=~ 5.7\times 10^{-15} \, \psi ~ a(q)^{-1} B^{-{q+1 \over 
2}} 250^{q-1 \over 2} ~~~ {\rm cm}^{-3} \,.
\label{eq:el_spect_norm}
\end{eqnarray}

A second relation is required to separately estimate $N_{\rm e}$ and $B$. 
This is provided by assuming that the energy density is equipartitioned 
between particles (electrons and protons) and the magnetic field, 
\begin{eqnarray}
U_{\rm p} + U_{\rm e} ~\simeq~ {B^2 \over 8\, \pi}\,.
\label{eq:equip}
\end{eqnarray}

The electron energy density is 
$U_{\rm e}$=$N_{\rm e,0}$$m_{\rm e}c^2$ $\int_{\gamma_1}^{\gamma_2} 
\gamma^{1-q} {\rm d}\gamma$, where $\gamma_1$$\simeq$100 corresponds to 
the energy below which Coulomb losses exceed 
synchrotron losses (the detailed number depends on the ambient, mostly 
thermal, density and magnetic field; e.g., 
Rephaeli 1979). For $q$$>$2 and $\gamma_2$$>>$$\gamma_1$, we get 
$U_{\rm e}$$\simeq$$N_{e,{\rm 0}} m_{\rm e}c^2 
\gamma_1^{2-q}/$$(q$$-$$2)$. Substituting the expression for $N_{{\rm e},0}$ 
from Eq.(\ref{eq:el_spect_norm}), 
we have 
\begin{eqnarray}
U_{\rm e} = 3.0\times 10^{-22} 
\, 250^{q \over 2}\, \psi \, {\gamma_1^{-q+2} 
\over (q-2)~a(q)}\, B^{-{q+1 \over 2}} \, {\rm erg~cm}^{-3}.
\label{eq:el_en_dens}
\end{eqnarray}
Equation (\ref{eq:el_en_dens}), together with Eqs.(\ref{eq:kappa_def}) and (\ref{eq:equip}), then leads to 
\begin{eqnarray}
\lefteqn{ B_{\rm eq} = 
\biggl[ 7.5\times 10^{-17}\,{ 
(2.5\times 10^{-2})^{q \over 2}
\over q-2}\, \psi \, {1+\kappa(q) \over a(q)} \biggr]^{2 \over 5+q}  {\rm G}. }
\label{eq:equip_B}
\end{eqnarray}

From the above relations it can be readily seen that if the flux, source 
size, distance, and the spectral index of the emitting electron population 
($q$) are known, then the source magnetic field -- and hence the particle 
energy density -- can be evaluated. For example, if $q$$=$2.5 (which 
corresponds to a typical galactic radio spectral index of 0.75, and 
entails $a$$=$0.0852 and $\kappa$$ \simeq$8) and $\psi =1$, we obtain 
$B_{\rm eq}$$=$61$\mu$G, 
$U_{\rm e}$$=$10.25 eV cm$^{-3}$, 
$U_{\rm p}$$=$82 eV cm$^{-3}$, 
$N_{\rm e,0}$$=$10$^{-4}\,$cm$^{-3}$.

\subsubsection{FIR radiation and Compton X-ray emission}

In intensely star-forming regions the stellar IMF is top-heavy (e.g., 
Mayya et al. 2007), so massive stars are proportionally more abundant 
there than in more typical Galactic environments. During most of their 
lifetimes massive stars are embedded in the dense, highly extinguished 
regions (giant molecular clouds) where they were born (e.g., Silva et 
al. 1998). The radiation emitted by these stars peaks in the UV, which 
is very efficiently absorbed by the embedding dusty clouds and is 
re-emitted in the FIR. Therefore the local radiation field is most 
intense in the FIR region, where it can be described as an isotropic, 
diluted, modified blackbody with temperature $T_{d}$ and spatial dilution 
factor $C_{\rm dil}$: 
\begin{eqnarray}
n_{\rm FIR} 
~=~ C_{\rm dil} ~ 
{1 \over \pi^2 (\hbar c)^3} ~ {\epsilon^2 \over e^{\epsilon/kT_{\rm d}} 
-1} ~ \bigl({\epsilon \over \epsilon_0}\bigr)^\sigma 
\label{eq:dil_BB}
\end{eqnarray}
with $0$$\leq$$\sigma$$\leq$$2$, for which we choose $\sigma$$=$1 (e.g., 
Goldshmidt \& Rephaeli 1995), and $\epsilon_0$ corresponding to 
$\nu$$=$2$ \times$10$^{12}$ Hz (Yun \& Carilli 2002, and references therein). 
For a given $T_{\rm d}$, the energy density of the warm dust, 
$U_{\rm FIR}(C_{\rm dil})$, can then be computed; comparison with the 
observed value, 
$U_ {\rm FIR}^{\rm obs}= L_{\rm FIR}^{\rm obs}$/$(\pi r_{\rm s}^2 c)$, 
where $r_{\rm s}$ is the radius of the accelerating source, yields 
$C_{\rm dil}$. 

When electrons whose energy distribution is as in Eq.(\ref{eq:el_spectrum}) 
interact with this FIR radiation, the spectral emissivity of the 
isotropically Compton-scattered radiation is (e.g., Rybicki \& Lightman 1979) 
\begin{eqnarray}
\lefteqn{
j_{\rm C}(\epsilon) ~ = ~ C_{\rm dil}~ F(q) ~ 
{N_{0,{\rm e}}  \, (e^2/mc^2) \over \pi^2 \hbar^3 c^2} ~ 
(kT_d)^{q+5 \over 2} ~ 
\epsilon^{-{q-1 \over 2}} } & 
\label{eq:IC}
\end{eqnarray}
where $F(q)$$\equiv$${q^2+4q+11 \over (q+3)^2 (q+5) (q+1)}$$\,\Gamma\bigl(
{q+5 \over 2}\bigr)$$\,\zeta\bigl({q+5 \over 2}\bigr)$, with $\Gamma$ 
and $\zeta$ denoting the Gamma function and the Riemann zeta function. 
For GeV electrons and typical dust temperatures, most of the emission is 
in the X-ray region.

\subsubsection{$\gamma$-ray emission from $\pi^{0}$ decay}

Proton energy losses at low energies are also dominated by Coulomb 
interactions with gas particles. At energies above pion masses 
($\sim$140 MeV), the main energy-loss process is by interactions with 
ambient protons; the yield from this process are neutral ($\pi^{0}$) and 
charged ($\pi^{\pm}$) pions. Neutral pions decay into photons, while the 
decays of $\pi^{\pm}$ yield relativistic e$^+$+e$^-$ and neutrinos. 

The integral spectral emissivity from $\pi^{0}$ decay is: 
\begin{eqnarray}
g_\eta(\geq 1\, {\rm TeV}) = g_\eta \bigl({\epsilon \over {\rm TeV}} 
\bigr)^{3-\eta} ~ \bigl({\rm erg \over cm^3}\bigr)^{-1} ~ ({\rm H~atom})^{-1} ~{\rm s}^{-1}
\label{eq:hadr_em}
\end{eqnarray}
(Drury et al. 1994), where $\eta$$=$$q$$+$2. This emissivity is very 
strongly dependent on the proton power-law index; e.g., for momentum 
distribution indices $\eta$$=$4.1, 4.3, 4.5, 4.7, we deduce (using 
tabulated values from Drury et al. 1994) that the $\geq$100 GeV emissivity, 
$g(\geq$100$\,$GeV), equals 
1.3$\times$10$^{-16}$, 4.2$\times$10$^{-17}$, 9.5$\times$10$^{-18}$, 
1.9$\times$10$^{-18}$ (erg cm$^{-3}$)$^{-1}$ 
(H~atom)$^{-1}$ s$^{-1}$, respectively.

The integrated $\gamma$-ray photon luminosity of a source with gas density 
$n$ and proton energy density $U_{\rm p}$ in a volume $V$ is then 
\begin{eqnarray}
L(\geq \epsilon) ~=~ \int_V g(\geq \epsilon) \,n\, U_{\rm p} 
\, dV ~~ {\rm s}^{-1}  & &
\label{eq:hadr_lum}
\end{eqnarray}
where both $g(\geq \epsilon)$ and $U_{\rm p}$ depend on the relativistic 
protons' spectral slope, $q$.

\section{VHE emission from M\,82: approximate treatment}

In this section we estimate the level of VHE emission from hadronic 
interactions in M\,82, the most prominent nearby ($d=3.6$ Mpc) 
starburst galaxy. The radio emission of M\,82 is observed to be power-law 
from $\nu_{\rm min}$$=$22.25 MHz to $\nu_{\rm max}$$=$41 GHz with a
flux of 10 Jy at 1 GHz, and the spectrum steepening from $\alpha \simeq 
0.71$ (Carlstrom \& Kronberg 1991) in the central starburst (defined here 
as a region with a radius of 300 pc and height of 200 pc; e.g., 
V\"olk et al. 1996, Mayya et al. 2006) to $\alpha \simeq 1$ (Klein et al. 
1988, Seaquist \& Odegard 1990) in the outer disk. The high-frequency radio 
data from the central disk region are shown in Figure 1. While there is 
some curvature in the radio spectrum, because we are mostly interested here 
in the high-energy end of the electron spectrum, this curvature is ignored 
and we fit the spectrum by a single power-law model. We infer that the 
synchrotron radio emission is from a population of relativistic electrons 
whose energy spectrum -- in the region sampled by the observed radio 
emission -- is a power-law with index $q$$\simeq$2.42.

Using the implied value of $q$, from Eqs.(\ref{eq:numratio2}) and 
(\ref{eq:kappa_for}) we obtain $\kappa$$\simeq$10 and $N_{\rm p}$$/$$N_{
\rm e}$$\bigr|_{1\, {\rm GeV}}$$\simeq$ 2$\times$10$^2$. Estimates of 
the Galactic values of these energy densities are $U_{\rm p}$$\sim$0.3 
eV cm$^{-3}$ and $U_{\rm e}$$\sim$0.03 eV cm$^{-3}$, hence 
$\kappa$$\sim$10 (see Schlickeiser 2002), and $N_{\rm p}$/$N_{\rm e}$$ 
\sim$10$^2$ at 1 GeV. The values of the two ratios are then quite similar 
in the two galaxies, in spite of their very different environments.

This similarity can be understood by the following argument. Typical SNR 
ages (hence, the duration of the acceleration processes), $\tau_{\rm SNR}$$
\mincir$2$\times$10$^4$ yr, are much shorter than the 
characteristic energy-loss timescales of both electrons and protons in 
the remnant (even for the strongest magnetic fields in SNRs allowed by 
current IACT data, $B= O(10^2) \,\mu$G; e.g., Berezhko \& V\"olk 2006), 
$\tau_{\rm s}$$\sim$1 Gyr and 0.5 Myr for GeV protons and electrons, 
respectively. Because of this, and the scaling of acceleration processes 
and propagation by rigidity, protons and electrons in the central disk 
region are expected to retain approximate similarity of their spectra at 
low to moderate energies (which dominate the particle distributions). 
Thus, the spectra have roughly the same spectral index both at injection 
($q_{\rm i}$$\simeq$2) and, after injection, while diffusing in the central 
disk with a diffusion coefficient $\propto$$p^b$ with $b$$\sim$0.5 
($q$$=$$q_{\rm i}$$+$$b$$\sim$2.5) (e.g., B\"usching et al. 2001). This 
will hold both in the very actively star-forming central region of M\,82 
and in the relatively quiescent Galactic environment.

Using $q$$=$2.42 (that entails $a$$\simeq$0.09 and 
$\kappa$$\equiv$$U_{\rm p}/U_{\rm e}$$\simeq$10), Eqs.(\ref{eq:equip_B}), 
(\ref{eq:el_spect_norm}), (\ref{eq:el_en_dens}), and the above-quoted 
parameters appropriate for M\,82 (which entail $\psi$$=$7.2), we derive: 
$B$$\simeq$106\,$\mu$G, $N_{\rm e,0}$$\simeq$1.4$\times$10$^{-4}$ cm$^{-3}$, 
$U_{\rm e}$$\simeq$25\,eV\,cm$^{-3}$, and $U_{\rm p}$$\simeq$250\,eV\,cm$^{-3}$.

\subsection{X-ray emission} 

Nonthermal X-ray emission is mostly by Compton scattering of the electrons 
by the FIR radiation field. To estimate the level of this emission, we first 
determine the energy density of the FIR field, whose dilution constant, 
$C_{\rm dil}$, which appears in Eq.(\ref{eq:dil_BB}), is evaluated as follows. 
Using $T_{\rm d}$$=$$45\,^{\circ}\mathrm{K}$ (Carlstrom \& Kronberg 1991) in 
Eq.(\ref{eq:dil_BB}), the warm dust energy density is 
$U_{\rm FIR}$$=$$5.55\,$$C_{\rm dil}\,$10$^{-8}$ erg cm$^{-3}$. On the 
other hand, the measured flux is $f_{\rm FIR}$$\simeq$6.5$\times$10$^{-8}$ 
erg cm$^{-2}$ s$^{-1}$, so that the observationally deduced value is 
$U_{\rm FIR}^{\rm obs}$$=$$L_{\rm FIR}^{\rm obs}$/$(\pi r_{\rm s}^2 c)$$
\simeq$1.26$\times$10$^{-9}$. Equating $U_{\rm IR}$ to $U_{\rm IR}^{\rm obs}$ 
yields $C_{\rm dil}$$=$0.023. (This value is slightly reduced if one corrects 
the observed FIR luminosity of M\,82 for cirrus emission.)

From Eq.(\ref{eq:IC}) the differential Compton flux from the central SB of 
M\,82 is $f_{\rm C}(\epsilon)$$\simeq$4$\times$10$^{-13}$$\epsilon^{-0.71}$ 
erg cm$^{-2}$ s$^{-1}$ keV$^{-1}$. 

The emission is clearly in the X-ray range; e.g., electrons with $\gamma$ as 
low as 150 and 350 boost a 60$\mu$m photon (representative of the FIR field) 
to, respectively, 2 keV and 10 keV. Given their steep energy distribution, 
only electrons of relatively low energy will be effective in generating a 
substantial Compton flux. Thus, electrons with 100$\leq$$\gamma$$\leq$1000 
scattering off the $45\,^{\circ}\mathrm{K}$ blackbody photons will boost 
their energies mostly to the 0.4-80 keV band. In particular, the predicted 
2-10 keV Compton flux is $f_{\rm C}(2-10\,{\rm keV})$$\simeq$10$^{-12}$ erg 
cm$^{-2}$ s$^{-1}$. 

There is yet no unequivocal evidence for nonthermal X-ray emission in M\,82. 
{\it Chandra} observations show that some of the 2-10 keV emission in M\,82 
is diffuse, emanating from a region which roughly overlaps with the central 
starburst (Griffiths et al. 2000). The 2-10 flux of this diffuse emission is 
$1.4\times 10^{-12}$ erg s$^{-1}$, some $5\%$ of the total meassured 2-10 keV 
emission (e.g., Cappi et al. 1999; Rephaeli \& Gruber 2002). Given that some 
of this diffuse component may be thermal, as suggested by the presence of a 
substantial 6.7 keV Fe-K emission, the observed diffuse emission is a strong 
upper limit on our predicted lower Compton contribution.

\subsection{VHE $\gamma$-ray emission}

The value $q$$=$2.4 deduced for the central SB region corresponds to 
$\eta$$=$4.4, for which we determine the appropriate value of $g$ from 
Table 1 of Drury et al. (1994), and obtain 
$g(\geq \epsilon)$$=$8.1$\times$10$^{-19} (\epsilon /{\rm TeV})^{-1.4}$ 
(erg cm$^{-3}$)$^{-1}$ atom$^{-1}$ s$^{-1}$, which holds down to GeV 
energies (e.g., Torres 2004; Domingo-Santamar{\`\i}a 2006).

A rough estimate of the VHE emission from M\,82 can be made if it assumed 
that this emission comes mostly from $\pi^0$ decay following interactions 
of energetic protons with ambient protons in the SB region. In this 
approximate treatment we also ignore contribution to the VHE emission by 
electron radiative processes. Assuming that acceleration occurs mostly in 
the central SB (of radius $r_{\rm s}$$=$300 pc), we thus take the proton 
energy density to be $U_{\rm p}(R) \sim 250 (R/r_{\rm s})^{-2}$ eV cm$^{-3}$ 
for $R$$>$$r_{\rm s}$, and consider separately the emission from the central 
SB and the disk:

\noindent
{\it (i)} In the central SB region ($R$$\leq$$r_{\rm s}$) the hydrogen mass 
is mostly molecular, $M_{\rm H_2}$$\simeq$2$\times$$10^8M_\odot$ (Weiss 
et al. 2001). The estimated VHE photon luminosity is then $L(\geq 100\, 
{\rm GeV})$$\simeq$2$\times$10$^{39}$ s$^{-1}$, which corresponds to 
$f(\geq 100\, {\rm GeV})$$\simeq$$1.3\times 10^{-12}$ cm$^{-2}$ s$^{-1}$. 

\noindent
{\it (ii)} The disk region external to the central SB, $R$$> $$r_{\rm s}$, 
has a flat thin disk gas distribution, $\Sigma(R)= \Sigma(0)\, e^{-R/R_D}$, 
with $\Sigma(0)$$\simeq$7.5$\times$10$^{22}$ cm$^{-2}$ the effective central 
density and $R_D$$\sim$0.82 kpc, the radial lengthscale (Mayya et al. 2006). 
The total gas mass, $M_{\rm gas}$$\simeq$2.5$\times$$10^9 M_\odot$, results 
from $M_{\rm HI}$$\simeq$0.7$\times$$10^9 M_\odot$ and $M_{\rm H_2}$$\simeq
$1.8$\times$$10^9 M_\odot$ (Casasola et al. 2004). The energetic proton energy 
density is assumed to have the form $U_{p}(R)$$=$200$\,(R/r_{\rm s})^{-2}$ 
eV cm$^{-3}$. From 
\begin{eqnarray}
L(\geq \epsilon) ~\simeq~ 2 \pi \, R_D^2 \,U_{p}\, 
\Sigma(0) \, g(\geq \epsilon) \int_{0.366}^{\infty} 
{e^{-x} \over x} \, dx ~~~ {\rm s}^{-1} \,, & &
\nonumber
\end{eqnarray}
we then obtain $L(\geq 100\, {\rm GeV})$$\simeq$1.9$\times$$10^{40}$ 
s$^{-1}$, that corresponds to $f(\geq 100\, {\rm GeV})$$\simeq$$1.2\times 
10^{-11}$ cm$^{-2}$ s$^{-1}$. 

This approximate treatment then yields a total photon luminosity, 
$\simeq$2.1$\times$10$^{40}$ s$^{-1}$, which translates to a total flux 
$f(\geq 100\, {\rm GeV}) \simeq 1.3\times 10^{-11}$ cm$^{-2}$ s$^{-1}$.

Before accepting the above estimate as physically meaningful, we need 
to ascertain that energetic protons do indeed leave their SN acceleration 
sites, and that VHE photons propagate freely through the galactic disk. 
Indeed, even in large, $R \sim 10$ pc, dense molecular clouds with 
typical ambient proton density of order $10^2$ cm$^{-3}$, the optical 
depth to pp interactions is very low, $\tau_{\rm pp}$$\sim$$\sigma_{\rm 
pp}$$n_{\rm p}$$l$$\sim$3$\times$10$^{-4}$, where the cross section for 
pp interactions is $\sigma_{\rm pp}$$\sim$$10^2$ mb$=$10$^{-25}$ cm$^{2}$ 
(e.g., Eidelman et al. 2004). This estimate is in obvious agreement with 
the fact that energetic protons at the relevant energy range are actually 
detected on Earth. 

Absorption of VHE photons is also negligible; the optical depth to pair 
production by interactions with 
ambient photons, $\tau_{\gamma \gamma}$$=$$n_{\gamma}$$\sigma_{\gamma 
\gamma}$$l$ (where $n_{\gamma}$ is the target photon number density, 
$\sigma_{\gamma \gamma}$ is the $\gamma \gamma$ cross section, and $l$ 
is the distance traveled), is also very small. For TeV photons, 
the peak cross section, $\sigma_{\gamma \gamma} =1.7 \times 10^{-25}$ 
cm$^2$, is for 
interaction with $\sim$2$\mu$m photons. The 2$\mu$m luminosity, 
$L_{2\mu{\rm m}}$$=$2$ \times$10$^{43}$ erg s$^{-1}$ (e.g., Silva et al. 
1998), implies a 2$\mu$m photon density of $\sim$0.5 cm$^{-3}$ (with 
a disk radius $r^{\rm gal}$$\sim$7 kpc), i.e. $\tau_{\gamma \gamma}$$
\sim$10$^{-3}$. We conclude that absorption of VHE photons 
in -- and indeed also along the l.o.s. to -- M\,82 is negligible.

\section{Particle and radiation spectra in M\,82: full numerical treatment}

As specified above, the central SB region (which will also referred to as 
the source region) with a radius of 300 pc and height of 200 pc (e.g., Volk 
et al. 1996), is identified as the main site of particle acceleration. 
The superposed particle spectra from all SN shock regions 
yields the power-law distributions with index $q$$=$$2$ (as mentioned in 
sect.2.1). The theoretically predicted $N_{\rm p}/N_{\rm e}$ ratio 
is more likely to be valid in this source region, as is also the assumption of 
equipartition. Accordingly, we infer $N_{\rm p}$ in the source region from 
$N_{\rm e}$, which itself is deduced from radio measurements in the central 
disk, as explained below.

\begin{figure}[h]
\label{fig:radio_spectrum}
\centering
\epsfig{file=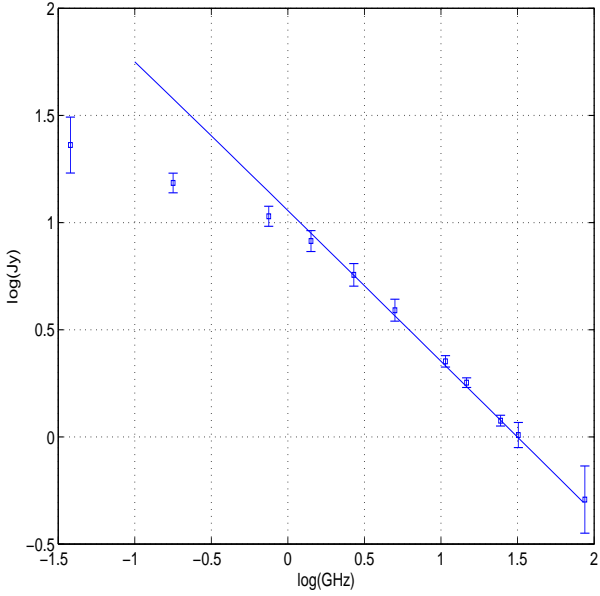,height=8cm,width=8cm,clip=}
\caption{Radio measurements (from Klein \ea 1988) and best-fit power-law 
spectrum to the data above $0.75$ GHz.}
\end{figure}

Particles are assumed to be convected out of their sources by the 
observed wind with velocity of 600 km $s^{-1}$ (Strickland et al. 1997). 
The convection velocity is assumed to increase linearly with distance from 
the disk plane. This assumption is consistent with cosmic-ray MHD wind 
models (e.g. Zirakashvili et al. 1996). The diffusion coefficient is taken 
as a power-law in rigidity $D_{xx} \propto (\rho/\rho_0)^\delta$, where 
$\rho_0$ is a scale rigidity with a typical index of 0.5. Particle energy 
losses and propagation outside the source region are followed in detail 
using the code of Arieli \& Rephaeli. To do so we need the neutral and 
ionized gas densities and their profile, as well as the spatial variation 
of the mean strength of the magnetic field. 

We use the neutral, molecular, ionized gas, and FIR radiation field 
parameters specified above. Additionally, we assume magnetic flux freezing 
in the (highly conductive) ionized gas, so that the local magnetic field 
strength is related to the ionized gas density, $n_{\rm HII}$, using the 
scaling $B$$\propto$$n_{\rm HII}^{2/3}$ (e.g., Rephaeli 1988). Measurements 
of radio emission from the central disk (Klein \ea 1988), shown in Figure 1 
together with our best fit power-law with index of $0.71 \pm 0.02$ (which is 
consistent with the results of Klein et al. 1988, and Carlstrom et al. 1993), 
provide the first relation between the electron density in this region and 
the magnetic field. Note that we ignore here the flattening indicated in the 
radio spectrum at $\nu$$\leq$300 MHz (which is possibly due to the increased 
relative strength of Coulomb losses at low electron energies) since we are 
primarily interested in the high energy behavior of the electron spectrum. 
The second required condition to fully determine particle spectra and the 
magnetic field is the assumption of equipartition in the source region. Due 
to the implicit dependences in the expression for the synchrotron flux, this 
condition is implemented iteratively to solve for $N_{\rm e}$, $N_{\rm p}$, 
and $B_0$. With the above parameters the central value of the magnetic field 
is $B_{0} \simeq 180$ $\mu$G. 

A radio index of $\sim$0.7 implies that the electron spectrum in the central 
disk steepens appreciably to $q_{\rm cd}$$\sim$2.4 (from $q=2$ in the source 
region). The steady-state electron and proton spectra in the SB region are 
shown in Fig.2. At low energies both spectra are flat; stronger electron losses 
at $E$$>>$1 GeV result in a steeper spectrum than that of protons, with the 
electron spectrum characterized by $q_{\rm cd}$$\sim$2.6 at very high energies.

\begin{figure}[h]
\label{fig:ep_density}
\centering
\epsfig{file=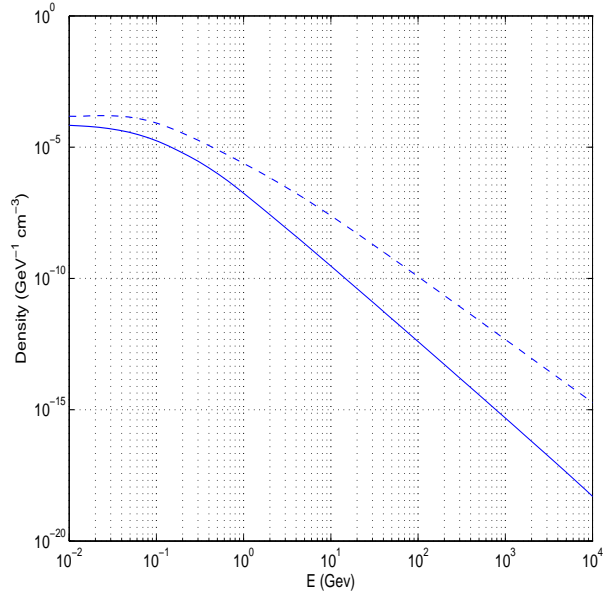,height=8cm,width=8cm,clip=}
\caption{Steady state primary electron (solid line) and proton  
(dashed line) spectral density distributions in the central disk 
region.}
\end{figure}

\begin{figure}[h]
\label{fig:spectra_all}
\centering
\epsfig{file=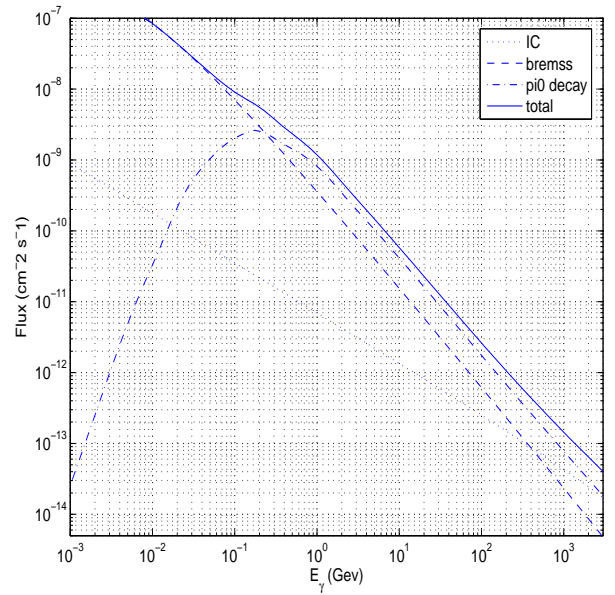,height=8cm,width=8cm,clip=}
\caption{Integral emission spectrum from the central disk region of M\,82. 
Radiative yields are from electron Compton scattering off the FIR radiation 
field (dotted line), electron bremmstrahlung off ambient protons (dashed 
line), $\pi^0$ decay following pp collisions (dashed-dotted line), and 
their sum (solid line).}
\end{figure}

Electron emissions by bremsstrahlung and Compton scattering are shown 
in Fig. 3; also shown is $\gamma$-ray emission from $\pi^0$ decay 
(following pp collisions). As expected, the losses due to bremsstrahlung 
dominate the lower energy regime, whereas losses due to $\pi^0$ decay 
dominate at higher energies. (While synchrotron emission extends to the 
X-ray region, it is negligible at much higher energies.) Our main interest 
here is the VHE emission, which -- as anticipated -- is mainly from the 
latter process. Specifically, we calculate the total flux from M\,82 
to be $f(\geq 100\, {\rm GeV})$$\simeq 2.1\times $10$^{-12}$ cm$^{-2}$ s$
^{-1}$. The flux at $\epsilon$$\geq$$50$ GeV is a factor of $\sim$3 higher.  

While we are mostly interested here in the VHE photon emission, we note 
that the related neutrino flux ($\pi^{\pm}$ eventually produce 
$e^{\pm}$ $+$ $\nu_{e}$ $+$ $\bar{\nu}_{\rm e}$ $+$ $\nu_{\mu}$ $+$ 
$\bar{\nu}_{\mu}$ ) at energies higher than 100 GeV is $\sim$0.3$\,
f($$\geq$100$\,{\rm GeV})$.

The predicted VHE $\gamma$-ray flux is a factor $\sim$6 lower than that 
estimated in the approximate treatment in sect.3.2. 
Given the nature of the approximations made, and the fact that spatial 
dependences of the particle densities and magnetic field were altogether 
ignored, this level of discrepancy is not surprising. 
The two treatments differ most in the description of the particle 
spatial profiles outside the source region. In the approximate 
treatment the impact of energy losses and propagation mode of the protons 
are not explicitly accounted for; this results in the unrealistically 
high relative contribution of the main disk (exterior to the SB region) 
to the TeV emission. Clearly, even a small degree of steepening of the 
proton spectrum in this region results in a significantly lower TeV 
emissivity (see, e.g., Table 1 of Drury et al. 1994) and hence flux.

\section{Discussion}

Our main objective has been an improved estimate of the VHE emission 
from the nearby northern-sky starburst galaxy M\,82. This has been obtained 
first by an approximate calculation of the hadronic emission alone, and then 
a more comprehensive and detailed treatment which follows the evolution of 
the steady state electron and proton spectra and their radiative yields, 
taking into account pertinent energy losses and propagation modes. (Note 
though that because VHE emission is largely produced in the central 
disk, details of our modeling of the diffusion coefficient and convection 
velocity are not that important in determining the particle spatial profile 
over this relatively small region.) Of course, only the full, more physically 
meaningful and realistic treatment provides a consistent basis for predicting 
the full particle and radiation spectra. The fact that we account for all 
known energy loss processes substantiates our expectation that the electron 
spectrum can be extrapolated to energies much higher than the range directly 
sampled by radio measurements. The approximate calculation is included here 
in order to give intuitive insight on the main contribution to the VHE 
emission, and in order to relate our work to previous similar treatments. 

As we have already noted, the factor of $\sim$6 
discrepancy between our two estimates of the total VHE flux is 
given the approximate nature of the analytic calculation. 
This is so mainly because the bulk of the VHE emission comes from a 
relatively small central region of M\,82 of $\sim$1 kpc radius, 
essentially the SB, due to steep decline of the emission outside this source 
region, which is described in the analytical treatment by $x^{-1}e^{-x}$. 
Over this region evolution of the particle spectra is moderate. Moreover, 
most of the VHE emission is from $\pi^0$ decay, hence a detailed 
calculation of the electron spectrum is not required. Of course, 
the approximate calculation is totally inadequate in predicting the lower 
energy emission, especially so over the larger disk region.

Compared to previous estimates of $\gamma$-ray emission from M\,82 (e.g., 
Aky\"uz et al. 1991; Pohl 1994; V\"olk et al. 1996), our detailed treatment 
is more realistic by virtue of being more tuned to a range of observables. 
These include the neutral and ionized gas densities and their estimated 
profiles across the central galactic disk, as well as the anticipated 
spatial variation of the magnetic field across this region. Most importantly, 
the basic normalization of the electron spectrum is directly determined from 
the radio spectrum measured in the central disk region. We assume 
particle-field energy equipartition and an initial, theoretically predicted 
$N_{\rm p}/N_{\rm e}$ ratio in the central 
region, where these are more likely to be valid, rather than across the full 
disk as in previous work. Finally, we include all the dominant radiative 
electron and proton processes. The calculation of Torres (2004), who 
estimated emission from the super-SB galaxy Arp\,220, and those of Paglione 
et al. (1996) and Domingo-Santamar{\`\i}a \& Torres (2005), who estimated 
VHE emission from the southern-sky starburst galaxy NGC\,253, although 
similar are still appreciably different from our numerical treatment. 
Given the substantial differences between our approach and those adopted 
in most previous treatments, it is not very meaningful to carry out a more 
detailed comparison of specific differences -- e.g., in parameter values and 
spatial profiles, including those of the electron and proton densities, and 
mean magnetic field in the central disk region -- and we avoid doing so here.

An assessment is needed of the reliability of our basic result -- the 
predicted level of VHE emission. Clearly, this flux depends linearly on 
$N_{\rm p}$, which in turn is related to $N_{\rm e}$. The electron density was 
deduced from the measured radio emission in the central disk region. 
Therefore, the evolution of the electron spectrum from the latter region to 
the central disk region is mostly determined by synchrotron losses in the 
high magnetic field. From Eqs. (\ref{eq:synchr}) and (\ref{eq:equip}) it 
follows that $N_{\rm p}$$\propto$$B_{eq}^2$ (which is obvious in the limit 
$U_{\rm p}$$>>$$U_{\rm e}$). The uncertainty in the estimated level of TeV 
emission stems mostly from this steep dependence. It is unlikely that the 
field is appreciably higher than our estimate ($B_0 \simeq 180$ $\mu$G), 
but it could possibly be lower. For example, a value lower by a factor of 
$\zeta$, would reduce $N_{\rm p}$ and $\pi^0$ decays by a factor $\zeta^2$. 
However, for a given measured radio flux $N_{\rm e}$ would have to be 
higher by the factor $\zeta^{(1+q)/2}$, which for a mean electron index 
of $\sim 2.4$ (over the central disk) would result in a boost of the 
bremsstrahlung and Compton yields by nearly the same factor ($\sim \zeta^2$). 
As can be seen in Figure 3, at the $\sim 100$ GeV region, the yield from the 
latter processes is only a factor $\sim 2$ lower than emission from $\pi^0$ 
decay, implying, e.g., that if the field were reduced to $B_0$$\simeq$100 
$\mu$G, the total VHE emission would not change significantly. 
Thus, the net effect of a lower field strength on the VHE emission is 
small. Of course, the $\pi^0$ decay yield is linearly dependent also on the 
ambient proton density, which we inferred directly from the measured mass 
in the central disk. We conclude that our predicted VHE flux is unlikely 
to be appreciably over-estimated. 
\footnote{By a numerical coincidence in spite of the different 
	    approach, our (numerical) VHE flux estimate is 
	    very similar to that by V\"olk et al. 1996, which was 
	    assumed to be emitted from the central starburst only, 
	    and to be purely hadronic with a hard ($q$$=$2.1) 
	    proton population throughout the emission region.}

Based on the above estimate, the predicted VHE flux of M\,82 
falls somewhat below the detection limit of current imaging air 
Cherenkov telescopes (IACTs). For example, MAGIC and 
VERITAS -- which are located in the Northern hemisphere and 
can therefore observe M\,82 -- have a 5$\sigma$ sensitivity of, 
respectively, $\sim$2$\times$10$^{-11}$ and 10$^{-11}$
cm$^{-2}$s$^{-1}$ for the detection of emission above 100 GeV in 50 
hours of observation (e.g., Bastieri et al. 2005). However, this 
sensitivity limit refers to a Crab-like spectrum, i.e. a power-law 
source whose differential VHE (photon) flux has an index of $2.6$, 
somewhat steeper than our deduced value of $\sim$2.3.
Accounting for this and for the inherent uncertainty (by a factor 
of $\mincir$2--3) in our lower flux estimate, we conclude that 
measurement of VHE emission from M\,82 -- or obtaining an interesting 
upper limit on it -- may be feasible with MAGIC or VERITAS, if 
observed for more than $\sim$500 hours. Prospects for detection are real 
with the upcoming MAGIC~II telescope, whose sensitivity at $\geq$100 GeV 
is expected to be a factor of $\sim$3 better than that of MAGIC I.

Finally, we comment on the possibility of detecting M\,82 at $\geq$100\,MeV 
$\gamma$-rays with {\it GLAST}. Our numerical code yields an integrated flux 
$f(\geq$100\, MeV)$\simeq$1.1$\times$10$^{-8}$ cm$^{-2}$s$^{-1}$. This value 
matches the 5$\sigma$ sensitivity of {\it GLAST}'s {\it Large Area Telescope} 
({\it LAT}) for a 1-yr scanning-mode operation (e.g., Dermer 2007). Thus, 
{\it GLAST/LAT} could possibly detect M\,82 during its first year of operation.

Its proximity and strong star-formation activity make M\,82 the most 
promising non-AGN galaxy target for high-energy $\gamma$-ray detection with 
the current generation of detectors ({\it GLAST/LAT}, MAGIC, and VERITAS). 
The LMC is the only non-AGN galaxy detected at energies $>$100\,MeV 
(Sreekumar et al. 1992); weak upper limits have been set on the VHE emission 
from normal and SB galaxies, including NGC\,253 (Aharonian et al. 2005c) and 
Arp\,220 (e.g., Albert et al. 2007). Detection of VHE $\gamma$-ray emission 
from nearby star-forming galaxies is clearly of much interest in view of the 
feasibility of testing cosmic-ray acceleration and propagation models. 
\medskip

{\it Acknowledgements.} We thank the referee for useful remarks 
on some aspects of particle acceleration by non-relativistic shocks.

\bigskip

\def\ref{\par\noindent\hangindent 20pt}

\noindent
{\bf References}
\vglue 0.2truecm

\ref{\small Aharonian, F.A., et al. 2007, A\&A, 464, 235 }
\ref{\small Aky\"uz, A., Brouillet, A., \& \"Ozel, M.E. 1991, A\&A, 248, 419}
\ref{\small Axford, W.I., Lear, E., \& Skadron, G. 1977, 
			in Proc. 15th ICRC (Plovdiv), 11, 132 }
\ref{\small Bastieri, D., et al. 2005, Astrop. Physics, 23, 572 }
\ref{\small Bell, A.R. 1978, MNRAS, 182, 443 }
\ref{\small Berezhko, E.G., \& V\"olk, H.J. 2006, A\&A, 451, 981}
\ref{\small Blandford, R.D., \& Ostriker, J.P. 1978, ApJ, 221, L29}
\ref{\small B\"usching, I., Pohl, M., \& Schlickeiser, R. 2001, A\&A, 377, 1056 }
\ref{\small Casasola, V., Bettoni, D., \& Galletta, G. 2004, A\&A, 422, 941 }
\ref{\small Cappi, M., Persic, M., Bassani, L., et al. 1999, A\&A, 350, 777}
\ref{\small Carlstrom, J.E., \& Kronberg, P.P. 1991, ApJ, 366, 422 }
\ref{\small Dermer, C.D. 2007, ApJ, 659, 958 } 
\ref{\small Domingo-Santamar\'\i a, E. 2006, PhD thesis (Universidad Aut\'onoma de Barcelona) }
\ref{\small Domingo-Santamar\'\i a, E., \& Torres, D.F. 2005, A\&A, 444, 403 }
\ref{\small Drury, L.O'C., Aharonian, F.A., \& V\"olk, H.J. 1994, A\&A, 287, 959 }
\ref{\small Goldshmidt, O., \& Rephaeli, Y. 2005, ApJ, 444, 113 }
\ref{\small Eidelman, S., et al. 2004, Phys Lett B, 592, 1 (see http://pdg.lbl.gov) }
\ref{\small Freedman, W.L., Hughes, S.M., Madore, B.F., et al. 1994, ApJ, 427, 628 }
\ref{\small Griffiths, R.E., Ptak, A., Feigelson, E.D., et al. 2000, Science, 290, 1325 }
\ref{\small Klein, U., Wielebinski, R., \& Morsi, H.W. 1988, A\&A, 190, 41 }
\ref{\small Krymsky, G.F. 1977, Dokl. Akad. Nauk SSSR, 243, 1306 }
\ref{\small Lerche, I., \& Schlickeiser, R. 1982, MNRAS, 201, 1041 }
\ref{\small Mayya, Y.D., Bressan, A., Carrasco, L., \& Hernandez-Martinez, L. 2006, ApJ, 649, 172 }
\ref{\small Paglione, T.A.D., Marscher, A.P., Jackson, J.M., \& Bertsch, D.L. 1996, ApJ, 460, 295 }
\ref{\small Moskalenko I.V, \& Strong A.W. 1998, ApJ, 493, 694 }
\ref{\small Moskalenko I.V, Jones, F.C., Mashnik, S.G., Ptuskin, V.S., \& Strong A.W., 2003, ICRC, 4, 1925 } 
\ref{\small Persic, M., \& De Angelis, A. 2008, A\&A, 483, 1 }
\ref{\small Persic, M., \& Rephaeli, Y. 2007, A\&A, 463, 481 }
\ref{\small Pohl, M. 1994, A\&A, 287, 453 }
\ref{\small Protheroe, R.J., \& Clay, R.W. 2004, PASA, 21, 1 }
\ref{\small Rephaeli, Y. 1979, ApJ, 227, 364 }
\ref{\small Rephaeli, Y. 1988, Comm. Ap., 12, 265 }
\ref{\small Rephaeli, Y., \& Gruber, D. 2002, A\&A, 389, 752 }
\ref{\small Rybicki, G.B., \& Lightman, A.P. 1979, Radiative Processes in Astrophysics, p.202--208 }
\ref{\small Rieke, G.H., et al. 1980, ApJ, 238, 24 }
\ref{\small Schlickeiser, R. 2002, Cosmic Ray Astrophysics (Berlin: Springer), p.472 }
\ref{\small Seaquist, E.R., \& Odegard, N. 1991, ApJ, 369, 320 }
\ref{\small Silva, L., et al. 1998, ApJ, 509, 103 }
\ref{\small Sreekumar, P., Bertsch, D.L., Dingus, B.L., et al. 1992, ApJ, 400, L67} 
\ref{\small Strickland, D.K., Ponman, T.J., \& Stevens, I.R. 1997, A\&A, 320, 378 }
\ref{\small Thompson, T.A., Quataert, E., \& Waxman, E. 2007, ApJ, 654, 219 }
\ref{\small Torres, D.F. 2004, ApJ, 617, 966 }
\ref{\small Tucker, W. 1975, Radiation Processes in Astrophysics (Cambridge, MA: MIT Press) }
\ref{\small Yun, M.S., \& Carilli, C.L. 2002, ApJ, 568, 88 }
\ref{\small V\"olk, H.J., Aharonian, F.A., \& Breitschwerdt, D. 1996, SSRv, 75, 279 }
\ref{\small V\"olk, H.J., Klein, U., \& Wielebinski, R. 1989, A\&A, 213, L12 }
\ref{\small Weiss, A., Neininger, N., H\"uttemeister, S., \& Klein, U. 2001, A\&A, 365, 571 }
\ref{\small Zirakashvili, V.N., Breitschwerdt, D., Ptuskin, V.S., \& Voelk, H.J. 1996, A\&A, 311, 113 }

\end{document}